\documentstyle[prb,aps]{revtex}
\begin{document}
\title{Quantum virial expansion approach to thermodynamics of $^4$He
adsorbates in carbon nanotube materials: Interacting Bose gas in one
dimension}

\author{Antonio \v{S}iber\thanks{E-mail: asiber@ifs.hr}}
\address{Institute of Physics, P.O. Box 304, 10001
Zagreb, Croatia 
}
\maketitle

\begin{abstract}
I demonstrate that $^4$He adsorbates in carbon nanotube materials can be
treated as one-dimensional interacting gas of spinless
bosons for temperatures below 8 K and for coverages such that all
the adsorbates are in the groove positions of the carbon nanotube
bundles. The effects of adsorbate-adsorbate interactions are studied
within the scheme of virial expansion approach. The theoretical
predictions for the specific heat of the interacting adsorbed gas are
given. \\
\begin{center}
{\bf Published in Phys. Rev. B 67, 165426 (2003)}
\end{center}
\end{abstract}


\begin{center}
I. INTRODUCTION \\
\end{center}

The adsorption of gases in nanotube-based materials has been recently a
subject of
considerable interest and many theoretical and experimental studies
focused on this phenomenon have been
reported. 
\cite{Teizer,Teizer2,Hallock,Migone,ColeCol,Boni1,ColePRL,Gatica1,Siber2,uptake,Siber1,Siber3}
The interest in the subject stems partially from
the possibility to use these materials as efficient gas
containers for hydrogen storage. \cite{H2store} Another cause of
the interest is that the nanotube materials provide a
very specific potential energy environment for the gas atoms
and molecules. In particular, the nature of this environment is such that
it reduces the effective dimensionality of adsorbates
\cite{Gatica1,Siber2} which at sufficiently low temperatures behave as a
one-dimensional (1D)
gas. This provides an excellent opportunity to study the interactions in
the 1D gas. The problem
of $N$ particles interacting mutually via binary interaction potentials in
one dimension has been thoroughly investigated
in the literature and there exist exact quantum and classical solutions
for very specific functional forms of the interaction
potential. \cite{Calogero,Lieb1,Lieb2,Girard} The aim of this
article is to investigate a realistic system in which the adsorbate
atoms (molecules) interact with a relatively complicated binary potential
that is attractive at large and repulsive at short interadsorbate
separations. \cite{Aziz}

The outline of this article is as follows. In Sec. II, the behavior of
single atom adsorbed on the surface of a bundle of single-wall
carbon nanotubes (SWCNT) is discussed. The range of temperatures in which 
the isolated adsorbates exhibit effective one-dimensional behavior is
discussed for $^4$He atoms. In Sec. III the calculation of second
virial coefficient for the interacting gas in 1D is briefly
outlined. A fully quantal approach is followed as the gas of interest is
composed of $^4$He atoms for which quantum effects are essential. In
Sec. IV the specific heat of adsorbed He gas is predicted and the
effects of He-He interactions are discussed. The results obtained are
compared with those for exactly solvable models \cite{Lieb1,Lieb2} and
qualitative agreement is found. The calculation of specific
heat in combination with experiments, some of them quite recently reported
\cite{Hone,Las1}, is expected to yield an additional insight in the 
thermodynamics of the adsorbed gas. In Sec. V, the influence of the
corrugation of the nanotube (which is neglected in Secs. II, III and
IV) and sample inhomogeneities on the thermodynamics of adsorbed gas is
thoroughly discussed. Sec. VI concludes the article and summarizes
the main results of the present work.

\begin{center}
II. BEHAVIOR OF ISOLATED $^4$He ATOMS ADSORBED ON THE SURFACE OF A BUNDLE
OF SWCNT's \\
\end{center}

Although there is still an ongoing discussion 
\cite{Teizer,Teizer2,Hallock,Migone,Siber2,uptake} concerning the
preferable adsorption sites for He atoms in SWCNT's materials, the
experimental information \cite{Teizer,Teizer2,Hallock} combined with the
theoretical considerations \cite{Siber2,uptake,Siber1} suggests that
individual He atoms are predominantly adsorbed in the groove positions on
the SWCNT's bundle surface. If the number of He atoms adsorbed in the
sample is very large, then one can expect that the He atoms will also
occupy other positions on the bundle surface. This point will be discussed
in Sec. III.

Quantum states of $^4$He atoms adsorbed in the groove positions of an
infinitely long bundle made of (10,10) SWCNT's have been discussed in
Ref. \onlinecite{Siber2}. It was found that the low-energy part of the
$^4$He excitation spectrum exhibits a typical 1D behavior with
characteristic $1/\sqrt{E}$ singularities present in the density of
states, $g(E)$. The density of states does not exhibit gaps which is a
consequence of the neglect of the corrugation of the carbon nanotube. In
this approximation there are no potential barriers for the adsorbate
motion along the groove. The severeness of this approximation and its
influence on the results to be presented shall be discussed in Sec. V.
In Fig. \ref{fig:fig1}, the low-energy part of
the $^4$He density of states per unit length of the groove is presented,
which was calculated as described in Ref. \onlinecite{Siber2}, i.e. the
three-dimensional Schr\"{o}dinger equation was numerically solved
to yield the complete set
of bound states. As can be seen in Fig. \ref{fig:fig1}, all the
excitations with energies between -22.7 meV and -18.88 meV pertain to
essentially one-dimensional $^4$He atoms. The
excitations in this regime of energies represent the activated, free
motion of $^4$He atoms
along the groove characterized by a 1D wave vector $K_y$ ($y$-axis is
oriented along the groove), as discussed in
Ref. \onlinecite{Siber2}. The transverse profile (in $xz$ plane,
perpendicular to the groove direction) of the $^4$He wave function is the
same for all these excitations and can be represented by a
narrow, Gaussian-like 2D function (see panel A of Fig. 4 in
Ref. \onlinecite{Siber2}).

At -18.88 meV, another band of states becomes available to the isolated
adsorbates. In this band, the transverse
profile of the $^4$He wave function is different from the ground-state
profile (see panel B of Fig. 4 in Ref. \onlinecite{Siber2}). As discussed
in Ref. \onlinecite{Siber2}, the population of higher (excited) bands
causes transition from the effectively 1D behavior of $^4$He atoms to 2D,
and eventually 3D behavior.

For the purposes of this work, it is sufficient to note that the
separation between the
lowest 1D band and the first excited band is quite large (3.82 meV) which
immediately suggest that the higher bands are
poorly populated in a significant range of temperatures. The width of 
this temperature range can be evaluated from the known density of
states. The total density of states can be represented as a sum of the
lowest band density of states, $g_0(E)$ and the density of states
representing all other transverse excitations, $g_{a}$,
\begin{equation}
g(E) = g_0 (E) + g_{a}(E).
\end{equation}
The lowest band density of states per unit length of the groove is given
by \cite{Siber2}
\begin{equation}
g_0 (E) = \sqrt{\frac{2m}{\hbar^2}} \frac{1}{2 \pi}
\frac{\Theta(E-E_{\Gamma})}{\sqrt{E-E_{\Gamma}}},
\end{equation}
where the mass of the adsorbate is $m$, $E_{\Gamma}=-22.7$ meV is
the ground state energy, and $\Theta$ is the Heaviside
function. The total number of adsorbates, $N$, is given by
\begin{equation}
N = L \int _{E_{\Gamma}} ^{\infty} g(E) f(E,T) dE,
\label{eq:Nnu}
\end{equation}
where $L$ is the total length of the groove, and the Bose-Einstein
distribution function is given by
\begin{equation}
f(E,T) = \frac{1}{\exp \left( \frac{E-\nu}{k_B T} \right) -1 }.
\end{equation}
Here, $k_B$ is the Boltzmann constant, $T$ is the temperature and $\nu$ is
the chemical potential. The number of
adsorbates in the lowest band, $N_{0}$, can be calculated as
\begin{equation}
N_{0} =  L \int _{E_{\Gamma}} ^{\infty} g_0(E) f(E,T) dE,
\end{equation}
once the chemical potential has been determined from Eq. (\ref{eq:Nnu}).

In Fig. \ref{fig:fig2}, I plot the ratio $N_0/N$ as a function of
temperature and for three different linear densities of adsorbates,
$n$, $n=N/L$. From this figure one can conclude that the {\em
noninteracting} $^4$He gas can be treated as effectively being 1D for
temperatures smaller than about 8 K (13 K) since for these temperatures
more than 99 \% (95 \%) of $^4$He atoms occupy the lowest energy band.

\begin{center}
III. VIRIAL EXPANSION APPROACH TO TREAT THE INTERACTING GAS IN 1D\\
\end{center}

The approach to be presented here assumes that all the adsorbed atoms are
in the groove positions on the bundle surface. Thus, the approach can
not be applied to the situations where the number of adsorbates is so
large that other positions on the bundle surface become occupied by the
adsorbates. One could envisage the situation where a very
large number of atoms is adsorbed on the bundle surface. From this phase,
one could get to the phase where all the adsorbates are exclusively in the
groove positions by desorbing all the atoms which are not in groove
positions. These atoms are more weakly bound than those in the
grooves\cite{Siber2} and will desorb from the sample at lower
temperatures. This fact enables the experimental realization of the
1D phase of interest to this work. Similar arguments can be applied to
1D heavy adsorbate phases studied in Ref. \onlinecite{Siber3}.

The virial expansion approach to treat the imperfect
(interacting) quantum
gas is well known \cite{Siddon,KHuang,Landsberg} and in this
section it will be only briefly outlined with reference to a quantum gas
in 1D.

The basic idea of the virial expansion approach is to represent the
so-called gas spreading pressure, $\phi$, as a power series of the gas
density, $n$,
\begin{equation}
\beta \phi = \sum _{l=1} ^{\infty} B_{l} (L, \beta) n^l.
\label{eq:spread}
\end{equation}
Here, $\beta = 1/k_B T$ and coefficients
$B_{l}$ are the virial coefficients. The virial coefficients can be
obtained by comparing the expansion of the gas spreading pressure in the
power series of fugacity, $z=\exp (\beta \nu)$,
\begin{equation}
\beta \phi = \frac{1}{L} \ln Q(z, \beta, L) = \sum_{l=1} ^{\infty} b_l (L,
\beta) z^l
\label{eq:pwrsz}
\end{equation}
with the expansion in Eq. (\ref{eq:spread}). $b_l$ is the $l$-th cluster
integral obtained as the coefficient in the power series expansion of the
logarithm of the grand partition function, $Q$, in terms of fugacity.
Since the grand partition function is given by
\begin{equation}
Q(z,L,\beta) = \sum_{l=0} ^{\infty} Z_l (L, \beta) z^l,
\label{eq:grandpar}
\end{equation}
where $Z_l$ are the quantum partition functions for $l$ particles, the
formulas for cluster integrals can be obtained by comparing the
expansions in Eqs. (\ref{eq:pwrsz}) and (\ref{eq:grandpar}). The
general form of $b_l$ is given in Ref. \onlinecite{Siddon}. For the first
two cluster integrals one has
\begin{eqnarray}
b_1 &=& \frac{Z_1}{L} \nonumber \\
b_2 &=& \frac{Z_2 - \frac{1}{2}Z_1^2}{L}.
\label{eq:bz}
\end{eqnarray}
Eliminating $z$ from equations (\ref{eq:spread}) and (\ref{eq:pwrsz}) by
expressing it in terms of linear density $n$, yields the relation between
the virial coefficients and cluster integrals. Using Eq. (\ref{eq:bz}), 
one can obtain the relations between virial coefficients and
quantum partition functions. Explicitly,
for the first and second virial coefficient one has
\begin{eqnarray}
B_1 &=& 1 \nonumber \\
B_2 &=& L \left ( \frac{1}{2}-\frac{Z_2}{Z_{1}^2} \right ),
\label{eq:vir1}
\end{eqnarray}

The quantum expression for the partition
function of $N$ particles in 1D is
\begin{equation}
Z_N = \int dy_1 ... dy_{N} \sum_{\alpha} \Psi_{\alpha}^* (y_1,
... ,y_n) \exp \left [ - \beta H(p_1, ... , p_N, y_1, ... , y_N) \right ]
\Psi_{\alpha} (y_1, ... ,y_n),
\label{eq:zn}
\end{equation}
where $p_i$ and $y_i$, $i=1,...,N$ represent 1D momenta and
coordinates of the $N$ gas particles, respectively. The dynamics of $N$ 
gas particles is described by the hamiltonian, $H(p_1, ... , p_N, y_1,
... , y_N)$. A complete set of quantum states describing $N$ particles
is denoted by $\{ \alpha \}$. The wave functions, $\Psi _{\alpha}$ are
assumed to be properly normalized and symmetrized according to the
statistics satisfied by the gas particles. It is easy to show
\cite{Siddon,KHuang,Landsberg} that the partition function for one 1D
spinless particle is given by
\begin{equation}
Z_1 = \frac{L}{\lambda},
\end{equation}
where $\lambda$ is the thermal wavelength, $\lambda = \sqrt{2 \pi \hbar
^2 \beta / m}$. Since I am going to consider $^4$He atoms, the
consideration of spinless particles will suffice. As shown by the
authors of Ref. \onlinecite{Siddon}, the spin degrees of
freedom can be considered (if needed) after the spinless problem has been
solved. In the problem of spinless particles (anti)symmetrization of the
wave function is
performed solely in the coordinate space. All expressions which follow do
not consider the spin degrees of freedom.

For the calculation of the second virial coefficient, $B_2$, given by
Eq. (\ref{eq:vir1}), one needs to calculate the partition function for two
interacting particles. This is an easy task when the corrugation of
the nanotubes can be neglected, since in that case the interacting 
two-body problem can be reduced to free motion of the center of mass
and the relative motion representing a particle of reduced mass $\mu =
m/2$ in an external potential \cite{Siddon,KHuang,Landsberg,Schif}. The
motion of the center of mass can be represented by a wave function for the
free particle of mass $2m$, $\exp (i k_{cm} Y) / \sqrt{L}$, where
$k_{cm}$ is the wave vector of the center of mass motion and $Y$ is the
center of mass coordinate, $Y=(y_1+y_2)/2$. The relative
motion can be described by a wave function of relative coordinate,
$y=y_1-y_2$, which is denoted by $\xi_{c}(y)$, and which satisfies 1D
Schr\"{o}dinger equation,
\begin{equation}
\left [ -\frac{\hbar ^2}{2 \mu} \frac{d^2}{dy^2} + v(|y|) \right ]
\xi_{c}(y) = \epsilon _{c} \xi_{c}(y).
\label{eq:xis}
\end{equation}
The binary potential representing an interaction between the two gas
particles is denoted by $v(|y|) = v (|y_1 - y_2|)$. The set of
"relative" quantum states is denoted by $\{ c \}$. This set consists of a
finite number of bound states denoted by $\{ b \}$, and the continuum of
states which can be numbered according to the wave vector $k$ associated 
with the motion of the particle with reduced mass in the region where the
interaction vanishes (large
$y$). The energy of quantum state $|c \rangle$ is denoted by
$\epsilon_c$. Relative wave functions behave in the asymptotic regime ($y
\rightarrow \infty$) as 
\begin{eqnarray}
\xi _{k} (y) &\rightarrow& \frac{1}{\sqrt{L}}\sin \left[ ky + \eta
(k) \right], \nonumber \\
\xi _{b} (y) &\rightarrow& 0
\end{eqnarray}
The 1D phase shifts, which are nonvanishing due to the presence of the
interaction potential $v$, are denoted by $\eta (k)$.

The partition function for the two {\em noninteracting} particles
[$v(|y|)=0$] in 1D can be calculated without invoking $Y$ and $y$
coordinates. Its form results solely from the requirement of the
(anti)symmetrization of the total wave function. Explicitly,
\begin{equation}
Z_2^{(0)} = \frac{Z_{1} ^2}{2} \pm \frac{L}{2\sqrt{2} \lambda},
\label{eq:z2non}
\end{equation}
where the upper(lower) sign is for spinless boson(fermion) gas. The second
virial
coefficient for the {\em noninteracting} gas is thus
\begin{equation}
B_{2}^{(0)} = \mp \frac{\lambda }{2 \sqrt{2}}.
\label{eq:b2non}
\end{equation}
The superscripts $(0)$ in Eqs. (\ref{eq:z2non}) and
(\ref{eq:b2non}) indicate that the expressions correspond to the
noninteracting quantum gas. As often noted in the literature
\cite{KHuang,Landsberg}, the finite value of $B_2^{(0)}$ coefficient for
noninteracting quantum gas reflects the so-called statistical attraction
for bosons (negative $B_2^{(0)}$) and statistical repulsion for fermions
(positive $B_2^{(0)}$, see Eq. (\ref{eq:spread})). The expression for
$B_{2}^{(0)}$ should be compared with Eq. (3.15) of
Ref. \onlinecite{Siddon} which pertains to noninteracting gas in two
dimensions. 

The second virial coefficient can be calculated as
\begin{equation}
B_2 = B_2^{(0)} + \lambda \sqrt{2} \sum_{c} \left[ \exp (-\beta
\epsilon_c^{(0)}) - \exp (-\beta \epsilon_c) \right],
\label{eq:B2temp}
\end{equation}
where $\epsilon_c^{(0)}$ is the set of "relative" energies for two
noninteracting particles. Eq. (\ref{eq:B2temp}) 
has been obtained by performing integration over the center of mass
coordinate in the expression for partition function of the two
interacting particles, $Z_2$ (Eq. (\ref{eq:zn})).

The formula for $B_2$ which is
convenient for numerical implementation can be obtained by replacing the
summation over $\{ c \}$ in
Eq. (\ref{eq:B2temp}) with two summations, one going over the bound
states, $\{ b \}$, and the other over the continuum states, $\{ k \}$. One
can pass from the sum over states $\{ k \}$ to the integral over wave
vector $k$ by introducing the density of
states in $k$ space, which can be obtained from the 1D phase
shifts, $\eta(k)$. \cite{Siddon} I finally obtain 
\begin{equation}
B_2(\beta) = B_2^{(0)} (\beta) - \lambda \sqrt{2} \sum_{b} \exp (-\beta
\epsilon_b) - \frac{\lambda \sqrt{2}}{\pi} \int_{0}^{\infty} \frac{d
\eta (k)}{d k} \exp \left( - \beta \frac{\hbar ^2 k ^2}{2 \mu} \right
) dk,
\label{eq:b2final}
\end{equation}
where the dependence of virial coefficient on the temperature (or
$\beta$) is emphasized. This formula is very similar to the one obtained
for 2D and 3D gas, although the numerical factors (such as $\sqrt{2}$) and
units of $B_2$ (in 1D, $B_2$ has units of length since the linear density
has units of inverse length) are different, depending on a
dimensionality of the problem. It should also be noted that since the
problem involves only one dimension, one does not obtain azimuthal quantum
numbers which occur in the treatments of interacting gas in 2D and 3D as a
consequence of the central symmetry of the binary potential.

The evaluation of Eq. (\ref{eq:b2final}) requires the calculation of
1D phase shifts which depend on the binary potential, $v$. The interaction
between the two He atoms in the otherwise empty space is known to a great
precision \cite{Aziz}. However, the effective interaction between the two
He atoms positioned in the vicinity of a third polarizable body is
different from the free space He-He interaction \cite{Bruchbook}. In the
case of interest to this work, the two He atoms are surrounded by two
SWCNT's and the polarization
induced in the SWNCT's will modify the He-He interaction. While the
polarization induced effects on the binary potential can be calculated for
atoms physisorbed on crystalline surface \cite{Bruchbook}, the analogous
calculation for the very specific geometry of the nanotube bundle is
certainly more difficult. It is interesting to note here, that
Vidali and Cole \cite{VidCol} found that the measurements of specific heat
of He overlayers on graphite \cite{Bretz} can be more accurately 
reproduced by the effective He-He potential which is 15 \% shallower from
the free-space He-He potential. They attribute this effect to the
screening of He-He interaction by the substrate. The treatment of Vidali
and Cole \cite{VidCol} was also based on quantum virial expansion. In
another study\cite{MKKost}, more related to the system considered here,
the authors found that the interaction between two He atoms adsorbed in
the interstitial channels of SWCNT's has a well depth which is 28 \%
shallower with respect to the free-space interaction. The groove
adsorption represents a situation which is "somewhere in between" the
adsorption on planar graphite and in SWCNT's interstitial channels.

In the following calculations, the He-He interaction will be described by
free-space potential suggested recently by Janzen and Aziz
\cite{Aziz}, but I shall also consider the scaled potential obtained from
the free-space interaction by simple multiplication with a
factor of 0.785. Thus, the scaled potential has a well depth which is 21.5
\% smaller from the well depth of the free-space potential. This number
was obtained as a simple arithmetic mean of the well depth reductions
found for adsorption on planar graphite\cite{VidCol} and in interstitial
channels of SWCNT's\cite{MKKost}. The assumed reduction of the well depth
is quite close to the numerical estimate in
Ref. \onlinecite{MKKost} (24 \%). The exact value of the scaling factor
used should not be taken too seriously because the substrate
induced contributions to the potential cannot be modelled by a simple
scaling of the free-space potential.\cite{Bruchbook} The scaled potential
was introduced simply to examine the effects of the details of interaction
potential on the thermodynamics of adsorbed gas. Additionally, to obtain
the effective potential in 1D, the 3D potential should be averaged over
the 2D cross-sections of the adsorbate probability density
\cite{VidCol,GuoBruch}. However, as
the cross-section of the lowest band states is rather small, and in the
light of the uncertainties of the substrate mediated forces, such a
procedure has not been performed.

SAPT1 potential supports one weakly bound state in 1D, representing a
$^4$He dimer with an energy of $\epsilon_0 = -0.16 \mu$eV. This state,
being so weakly bound, is extremely extended in relative
coordinate. \cite{Vranjes1,Vranjes2}. The bound state energy in 1D is
significantly smaller from the one obtained by Siddon and Schick in a 2D
treatment\cite{Siddon}, which is in accord with existing 
literature\cite{Vranjes1}. The scaled SAPT1
potential does not support bound states. The 1D phase
shifts and their derivatives with respect to relative wave vector were
calculated by numerically solving the Schr\"{o}dinger equation using the
algorithm\cite{privsib} quite similar to the one described in
Ref. \onlinecite{Gordon}.

In Fig. \ref{fig:fig3}, the calculated values of second virial coefficient
are presented. The full line corresponds to calculation with the
He-He potential suggested in Ref. \onlinecite{Aziz} (SAPT1), while the
dashed line corresponds to calculation using the scaled SAPT1
potential. 
The behavior of the second virial coefficient with temperature is
qualitatively similar to the one obtained for $^4$He adsorbates on
graphite (2D problem) in Ref. \onlinecite{Siddon}. There is, however, one
important difference. The ideal gas term, $B_2^{0}$, reflecting the purely
quantum effect of gas statistics, decays with
temperature as $\sqrt{1/T}$ in 1D, and as $1/T$ in
2D.\cite{Siddon}. Thus, the approach of second virial coefficient to its
classical value as the temperature increases is slower in 1D than in
2D. In the inset of
Fig. \ref{fig:fig3} the derivatives of the phase-shifts, $d
\eta (k) / d k$ are plotted as a function of relative wave vector
$k$. Note that the phase-shift derivatives become negative and nearly
constant for large relative energies (wave vectors). This is a consequence
of a strongly repulsive potential at short distances (hard core).
Note also that the phase shifts of the two potentials are very different
for small wave vectors, and thus one could expect that the two potentials
produce quite different thermodynamical quantities. However, the very
different behavior of the phase shifts is a consequence of the fact that
the SAPT1 potential supports a weakly bound state whereas the scaled SAPT1
potential does not. Thus,
in the evaluation of Eq. (\ref{eq:b2final}) one has to properly account
for the bound state which exists in the case of SAPT1 potential. This
"extra" term for SAPT1 potential makes the thermodynamical quantities
derived from the two potentials quite similar, although the derivatives of
the phase shifts are very different. This fact has been discussed for
the interacting gas in 2D \cite{Portnoi,Lin} in connection with Levinson's
theorem \cite{Schif,Levinson} which relates the phase shift at zero
momentum to
the number of bound states. It was found \cite{Portnoi} that a proper
account of both the continuum and bound states eliminates discontinuities
in thermodynamic properties whenever an extra bound state appears with the
small change of the parameters of interaction potential. In the present
calculation similar effect is found in one dimension.

\begin{center}
IV. SPECIFIC HEAT OF ADSORBED $^4$He GAS\\
\end{center}

The specific heat of interacting quantum gas can be calculated from
the set of virial coefficients \cite{KHuang,Landsberg}. I assume that the
dominant contribution to the specific heat comes from the second virial
coefficient. Thus, the results are applicable to a restricted range of
adsorbate concentrations and temperatures. The range of a validity of this
approximation can be estimated from a calculation of higher virial
coefficients, which is a difficult task, or from direct comparison with
experiments, as has been done for $^4$He adsorbates on graphite in
Refs. \onlinecite{Siddon} and \onlinecite{Elgin}. Experiments dealing
primarily with the specific heat of adsorbates in carbon nanotube
materials have not been reported yet and those which detected the
signature of the adsorbed gas in the overall specific heat of the sample
were focused on the specific heat of clean nanotube materials
\cite{Hone,Las1}.

The isosteric specific heat is given as \cite{Siddon}
\begin{equation}
\frac{C}{Nk_B} = \frac{1}{2} - n \beta^2 \frac{d^2 B_2}{d
\beta^2},
\label{eq:spechea1}
\end{equation}
The second term in Eq. (\ref{eq:spechea1}) can
be calculated from Eq. (\ref{eq:b2final}) as
\begin{equation}
n \beta^2 \frac{d^2 B_2}{d \beta^2} = n \lambda \sqrt{2} \left [ 
\frac{1}{16} + \frac{S_0 + I_0}{4} 
+ \beta \left ( S_1 + I_1 \right ) - \beta ^2 \left(S_2 + I_2
\right) \right ],
\label{eq:term}
\end{equation}
where
\begin{equation}
S_n = \sum_b \epsilon_b^n \exp (-\beta \epsilon_b),
\label{eq:sns}
\end{equation}
and
\begin{equation}
I_n = \frac{1}{\pi} \left ( \frac{\hbar^2}{2 \mu} \right ) ^n \int_0
^{\infty} k^{2n} \frac{d \eta (k)}{dk} \exp \left (-\frac{\hbar^2 k^2
\beta}{2 \mu} \right) dk.
\label{eq:ins}
\end{equation}
The expression for the term in Eq. (\ref{eq:term}) representing a
deviation of the specific heat from its ideal value (where the
interadsorbate interactions are neglected) is different from the
corresponding expression one would obtain in a 2D treatment. The 2D
expression\cite{Siddon} does not contain a term proportional to $\lambda$
and independent of the interadsorbate interaction potential (1/16 in
Eq. (\ref{eq:term})). The reason for this is that in 2D, the
noninteracting value of $B_2$, $B_2^{(0)}$ is proportional to $\beta$
($B_2^{(0)}=-\lambda^2 / 4$) \cite{Siddon}, and
therefore, its second derivative with respect to $\beta$ vanishes. In 1D
case, $B_2^{(0)} \propto \sqrt{\beta}$ (see Eq. \ref{eq:b2non}) and $d^2
B_2^{(0)} / d \beta^2 \propto \beta ^{-3/2}$. This fact alone suggests
that the specific heats of dilute, interacting boson gas in 1D and 2D may
be qualitatively different.

In Fig. \ref{fig:fig4} I plot the quantity $-\beta^2 d^2 B_2 / d
\beta^2$. The full (dashed) line represents the calculation
with SAPT1 (scaled SAPT1) potential. Obviously, both potentials produce
similar deviations. The calculation with scaled SAPT1 potential yields
somewhat smaller effects of interactions on the specific heat, which
is plausible since the scaled SAPT1 potential is weaker than
SAPT1 potential. An
important observation is that the deviation of specific heat from its
ideal value is {\em negative} for 1D boson gas, i.e. the inclusion of
interactions reduces the specific heat of the adsorbed 1D gas. For 2D
spinless boson gas considered in Ref. \onlinecite{Siddon}, the deviation
was found to be {\em positive}. The
observed difference between 1D and 2D results is a consequence of both the
nonvanishing second derivative of the ideal term
[$B_2^{(0)}(\beta)$] with respect to $\beta$, and
nonexistence of azimuthal degrees of freedom in 1D treatment of the
problem.

Finally, in Fig. \ref{fig:fig5}, I plot the specific heat of the
interacting $^4$He gas adsorbed in grooves of SWCNT bundles for three
different linear densities. The full (dashed) lines represent the
calculation with SAPT1 (scaled SAPT1) potential. Obviously, for denser
gas, the specific heat is more strongly influenced by the
interactions, and deviates more from the ideal (noninteracting) 1D value
(compare with Fig. 6 of Ref. \onlinecite{Siber2}). It can also be inferred
from Fig. \ref{fig:fig5} that the description of thermodynamics of the
adsorbed gas in terms of second virial coefficient only breaks down at
very low temperatures where such an approach yields negative specific
heat. These
temperatures can be considered as the lower limits for the application of
the present approach. For higher densities, the lower temperature limits
obviously increase, in agreement with Eq. (\ref{eq:spechea1}), and the
presented virial expansion approach breaks down at higher
temperatures. Note that the upper temperature for which the specific heat
was calculated is 6 K. Above that temperature the higher bands start 
to contribute to the specific heat as discussed in Sec. II and
Ref. \onlinecite{Siber2}. Although the
number of particles in higher bands is very small at 6 K (see
Fig. \ref{fig:fig1}), the derivative of internal energy
with respect to temperature (specific heat) is very sensitive even to
very small occupations of the higher bands and this causes the
increase of the specific heat observed for a very dilute gas.

It is now of interest to compare the obtained results with those of
exactly solvable 1D many-particle models. In particular, for the gas of
bosons in one dimension interacting via repulsive $\delta$-function
potential, Lieb and Liniger have demonstrated that the that the energy
spectrum of such a gas is identical with the spectrum of noninteracting
Fermi gas.\cite{Lieb1,Lieb2} The same was found in the model of
Girardeau\cite{Girard} for
1D bosons interacting with the binary hard-core potentials of finite
radius. The Fermi energy, $E_F$ of the corresponding 1D Fermi gas is given
as
\begin{equation}
E_F = \frac{\hbar^2 k_F^2}{2m} = \frac{\hbar^2 \pi^2}{2m} n^2,
\label{eq:ef}
\end{equation}
where $k_F=\pi N/L =\pi n$ is the Fermi wave vector. This mapping
(interacting Bose gas - noninteracting Fermi gas) allows us to easily
predict the specific heat of the impenetrable 1D Bose gas. Thus, at 
temperatures much smaller than the Fermi temperature, $T_F=E_F /k_B$,
\begin{equation}
C_V(T \ll T_F) \propto T,
\label{eq:linc}
\end{equation}
while at temperatures much larger than the Fermi temperature, specific
heat reduces to its classical equipartition value, 
\begin{equation}
\frac{C_V (T \gg T_F)}{N k_B} \rightarrow \frac{1}{2}.
\label{eq:proc}
\end{equation}
The linear temperature dependence of specific heat [Eq. (\ref{eq:linc})]
can be also interpreted as a signature of long-wavelength compression
waves of the 1D Bose gas (sound), and is thus also expected in a gas of
1D bosons interacting with more complex forces. This is a collective
effect and is obviously outside of the scope of the second order virial
expansion, treating only the two-body collisions. On the other hand, the
classical equipartition value of specific heat [Eq. (\ref{eq:proc})] is
obtained also by the present approach at high temperatures
[Eq. (\ref{eq:spechea1}) and Fig. \ref{fig:fig4}]. The temperature
separating
the two characteristic behaviors in Eqs. (\ref{eq:linc}) and
(\ref{eq:proc}) is roughly given by the Fermi temperature of the
equivalent noninteracting Fermi gas, which is proportional to the square
of linear density [Eq.(\ref{eq:ef})]. Thus, for denser Bose gases, the
interval of temperatures in which the specific heat is significantly
smaller than 1/2 is larger. This is again in agreement with
the results presented in Fig. \ref{fig:fig5}.

\begin{center}
V. THE EFFECT OF THE CORRUGATION OF CARBON NANOTUBE AND SAMPLE
INHOMOGENEITIES
\end{center}

The results presented in this article are based on an idealized
representation of the SWCNT materials. In particular, it is assumed that
SWCNT bundles are infinitely long, straight and smooth which is certainly
not the case in real materials and the bundles wiggle on large length
scales. The discreteness of the nanotube results in the corrugation of the
potential experienced by adsorbates which is not accounted for in the
present approach. 
The influence of the corrugation of the potential confining  
the He atom to the interstitial channel of the bundle
composed of (10,10) carbon nanotubes was
studied in Ref. \onlinecite{Siber1}. A model potential
was used which enabled an easy examination of effects
induced by the potential details. Substantial effect
of the corrugation on the density of states of single
He atom was predicted. In particular, for the lowest
band of states, a mass enhancement factor of 2.37 was calculated.
Another study \cite{Boni1} predicted a mass enhancement
factor of 1.3. A large difference between these factors   
predicted by the two studies can be attributed to the larger
separation of the tubes (by 0.1 \AA) adopted in Ref.
\onlinecite{Boni1} with respect to Ref. \onlinecite{Siber1}. 
For comparison, the mass enhancement factor for $^4$He   
adsorbed on graphite was found to be only 1.06. \cite{VidCol}
For an interstitial channel surrounded by three
(17,0) nanotubes, the authors of Ref. \onlinecite{Boni1}
found that a change in the intertube separation by
0.1 \AA \hspace{0.7mm} changes the effective mass of $^4$He by
a factor of two!

It is quite remarkable that a small change in the intertube
separation results in a very large change of the band structure
of adsorbates. A similar sensitivity of the adsorbate bound
states on the interaction potential was
found in Ref. \onlinecite{Siber2} for the groove
adsorption. The details of the interaction potential are
"magnified" in the spatially restricted regions of the
interstitial channel and the groove, and the band structure
of adsorbates is much more affected by the potential details when
compared with the adsorption on planar graphite. Thus, in
order to assess the effect of the corrugation on the
calculations presented in this article, one would need
to know the intertube separation distance to a precision
better than at least 0.05 \AA. Note that the groove region
is at the surface of the bundle where the relaxation effects  
can be expected and the separation between the tubes surrounding
the groove needs not to be the same as the separation between
the two tubes in the interior of the bundle.
Additionally, one would have to know
the He - single tube potential with a great accuracy.
Even then, the calculation of the corrugation of
the potential would require knowledge of the alignment
of the two tubes surrounding the groove. It is very likely
that this arrangement is not the same for all the grooves
on the bundle surface. Furthermore, if the two tubes
surrounding a groove are not of the same symmetry, the
corrugations of the two tubes are not necessarily
commensurate and the total potential for the groove 
adsorption can not be written as a Fourier series.

All the mentioned complications are not present for
the adsorption on planar graphite. In that case, a
fairly reliable potential can be constructed
\cite{Colerevmod} and written as a 2D
Fouries series. However, even the inclusion of
periodic corrugation in the formalism of
quantum virial expansion is not straightforward.
The basic reason for this is that the
two body Schr\"{o}dinger equation in the
textured potential background no longer separates
into the two equations describing the motion of
the center of mass and the relative motion.
For the adsorption of He on graphite, Guo and
Bruch \cite{GuoBruch} devised a perturbation
treatment in which the two-body cluster integral    
is written as an expansion in powers of the
Fourier amplitudes of the atom-substrate potential.
They also presented the results for the second
virial coefficient in which the corrugation effects
were treated up to the second order of the perturbation
series. Another study dealing with the problem
\cite{Rehr} started from the tight-binding
Hamiltonian and He atoms localized on particular
adsorption sites. A series of approximations
and simplifications was needed to obtain the
formula for the second virial coefficient
which is amenable to evaluation.

It is clear from the discussion in this section
that a reliable calculation of the second
virial coefficient with the effects of the corrugation
included is not possible at present, mainly because 
the relevant potential is not known with sufficient
precision. Additionally, aperiodic corrugation has 
not been treated in the literature in this
context. However, a simple estimate of the
corrugation effects is possible if we represent the
adsorbate as a particle with the effective mass. Such
a representation of the corrugation effects is
adequate for excitations with small velocity i.e.
for the states located around the center of the
1D Brillouin zone. The mass enhancement for the
groove adsorption can be expected to be somewhere
in between 1.06 (planar graphite) and 2.37
(interstitial channel). The quantity
$-\beta^2 d^2 B_2 / d \beta ^2$ was calculated using the
Eqs. (\ref{eq:term}), (\ref{eq:sns}) and (\ref{eq:ins}), 
the scaled SAPT1 potential, and the
effective He mass equal to $M^*=1.3 M_{He}$=5.2 amu.  
At $T=$1K (5K), its value was found to be -1.69 \AA
(-0.52 \AA). This should be compared with the values obtained
in Fig. \ref{fig:fig4}. In particular, at
$T=$1K (5K) the values obtained using the 
free mass of $^4$He are -1.92 \AA (-0.77 \AA). Thus,
the reflection of the corrugation effects on the
specific heat can be significant, but the overall
trends, at least in the model of renormalized
mass, are the same. For strongly corrugated potentials, 
it may be more sensible to use the band width for the 
characterization of the effects of corrugation rather than the 
curvature of dispersion curves at small wave vectors.\cite{Dash}

The influence of other bundles on
the atoms adsorbed in a groove of a particular bundle is neglected. This
approximation obviously breaks down if two bundles touch each other. At
these points, the adsorbates move in a potential very much influenced by
both bundles in question, which may be significantly different from
the potential of a single, infinitely long and straight bundle I used in
the calculations. For the adsorption of $^4$He on
graphite\cite{Bretz,Elgin}, it was argued
\cite{Campbell} that the presence of long-range inhomogeneities in the
graphite may act as
a trap and induce the Bose-Einstein condensation of the adsorbed gas. The
same experiments were later explained in terms of the virial expansion
approach \cite{Siddon} by assuming a perfect graphite
substrate. Nevertheless, a possible substantial influence of the
sample inhomogeneities on the thermodynamics of the adsorbed gas cannot be
{\em a priori} ruled out. This is expected to be more important at low
temperatures where the adsorbate thermal wavelength is large and the long
range order of the sample may influence the adsorbate gas
thermodynamics. It is hoped that the low-temperature measurements of the
specific heat may give answers to these questions.

\begin{center}
VI. SUMMARY AND CONCLUSION\\
\end{center}

$^4$He gas adsorbed in the grooves of single
wall carbon nanotube bundles has been treated as an interacting Bose gas
in one dimension. It was found that this approximation should be
very accurate for all temperatures below 8 K. The interactions
in the adsorbed gas are treated via the quantum virial expansion approach
and the second virial coefficient for the interacting gas was
calculated. This information was used to calculate the specific heat of
adsorbed $^4$He gas which was shown to be substantially influenced by
interadsorbate interactions already at relatively low adsorbate linear
densities (0.033 1/\AA). A qualitative agreement between the results
obtained in this article and those of exactly solvable 1D models is
demonstrated.

\begin{center}
ACKNOWLEDGMENTS\\
\end{center}

It is a pleasure to thank D. Jurman for stimulating discussions concerning
exactly solvable 1D many-particle models and for suggesting
Ref. \onlinecite{Calogero}.

\begin{figure}
\caption{
Density of states per unit length of the groove
of single $^4$He atom adsorbed in the groove of a bundle made of
(10,10) SWCNT's. The low energy portion of the density of states is
displayed. 
}
\label{fig:fig1}
\end{figure}

\begin{figure}
\caption{
Ratio of the number of $^4$He atoms in the lowest 1D band and the total
number of $^4$He atoms ($N_0/N$) as a function of temperature and for
three different linear densities. Thick full line: $N/L$=0.01 1/\AA. Thick 
dashed line: $N/L$=0.1 1/\AA. Thick dotted line: $N/L$=0.2 1/\AA. Two thin
dotted lines represent the 0.99 and 0.95 values of the ratio.
}
\label{fig:fig2}
\end{figure}

\begin{figure}
\caption{
Second virial coefficient for 1D $^4$He gas as a function of
temperature. Full line: Quantum calculation with SAPT1 potential. Dashed
line: Quantum calculation with scaled SAPT1 potential (see
text). Inset: Phase shift derivatives, $d \eta (k) / d k$, corresponding
to SAPT1 potential (full line) and scaled SAPT1 potential (dashed line).} 
\label{fig:fig3}
\end{figure}

\begin{figure}
\caption{
Deviation of the specific heat per unit linear density 
($-\beta^2 d^2 B_2 / d \beta^2$) from its ideal
value, Eq. (\ref{eq:spechea1}). Full line: Calculation with
SAPT1 potential. Dashed line: Calculation with scaled SAPT1 potential.} 
\label{fig:fig4}
\end{figure}

\begin{figure}
\caption{
Specific heat of $^4$He gas adsorbed in the grooves of SWCNT bundles for
three different linear densities (0.033 1/\AA, 0.1 1/\AA, and 0.2 1/\AA).
Full line: Calculation with
SAPT1 potential. Dashed line: Calculation with scaled SAPT1 potential.} 
\label{fig:fig5}
\end{figure}


\begin{references}
\bibitem{privsib} FORTRAN 77 code for the calculation of phase shifts can
be obtained from the author.
\end{references}
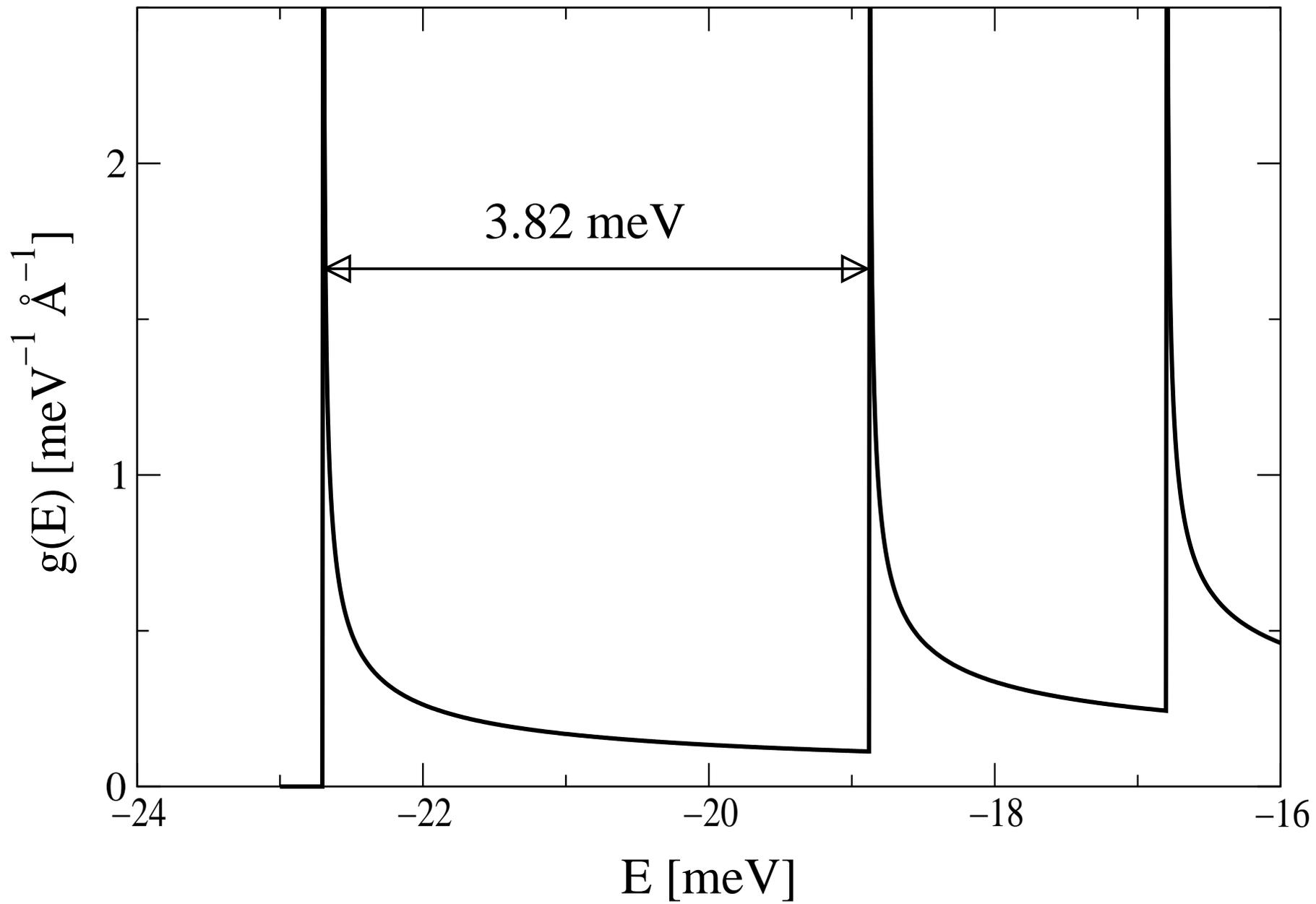
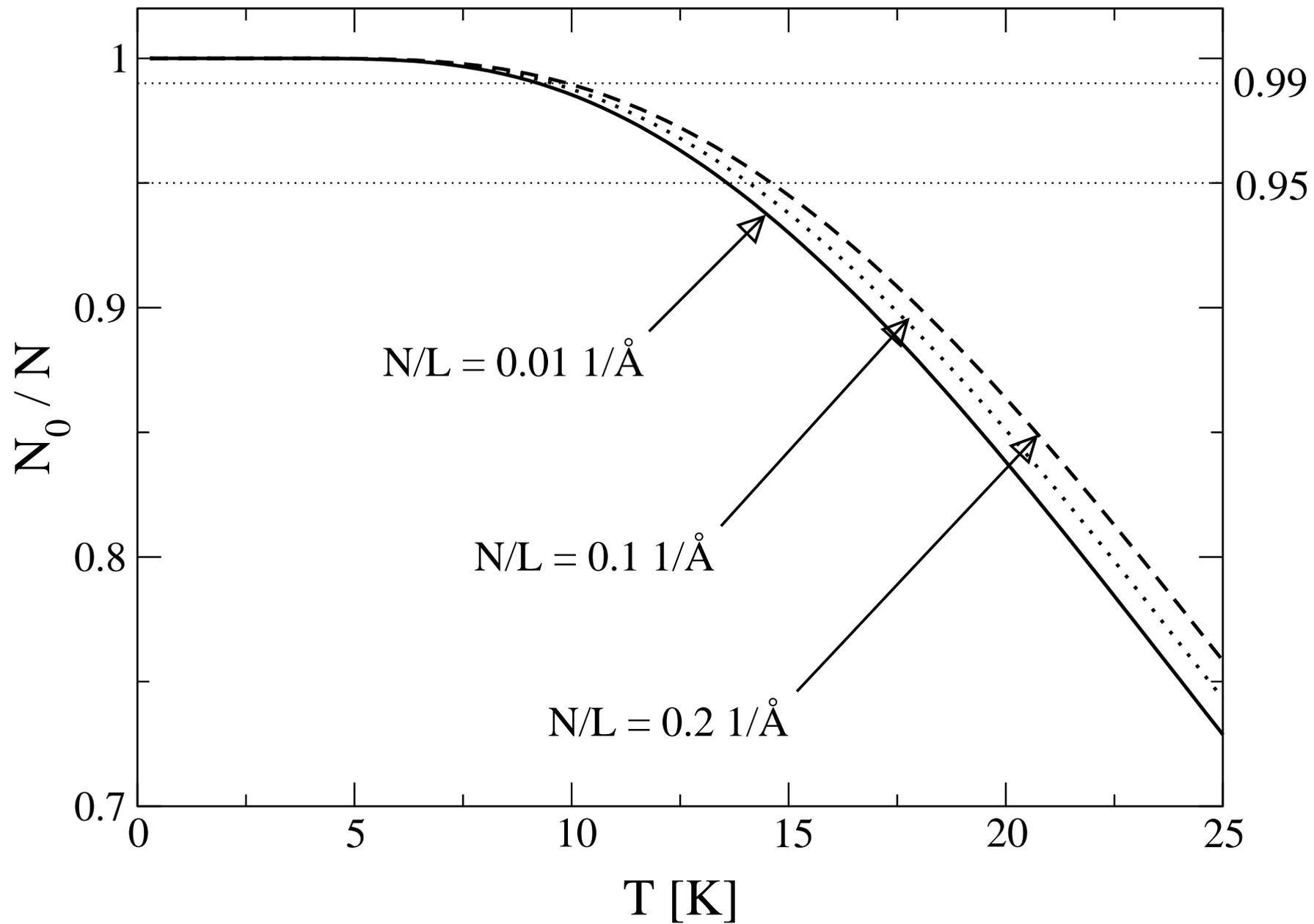
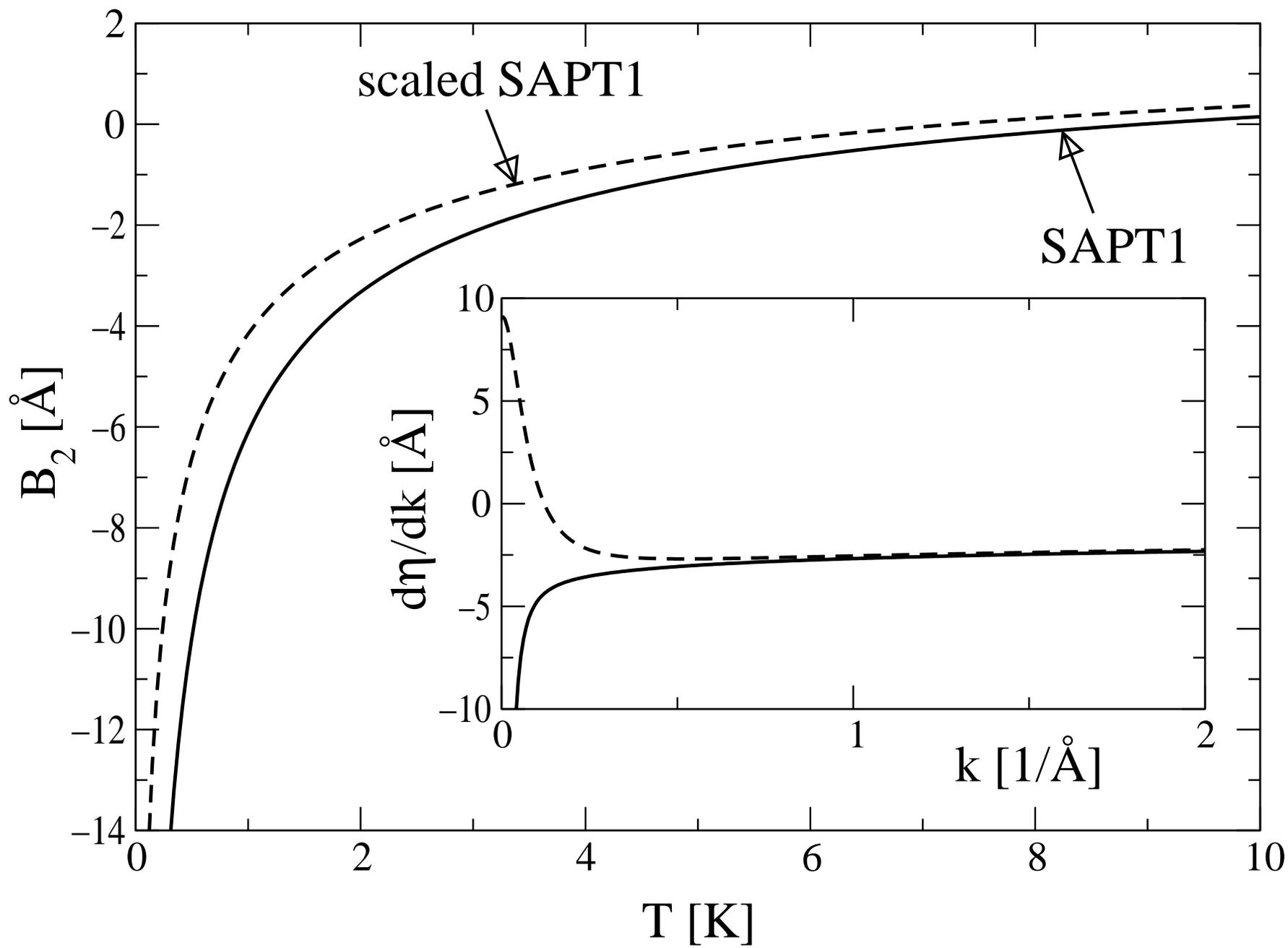
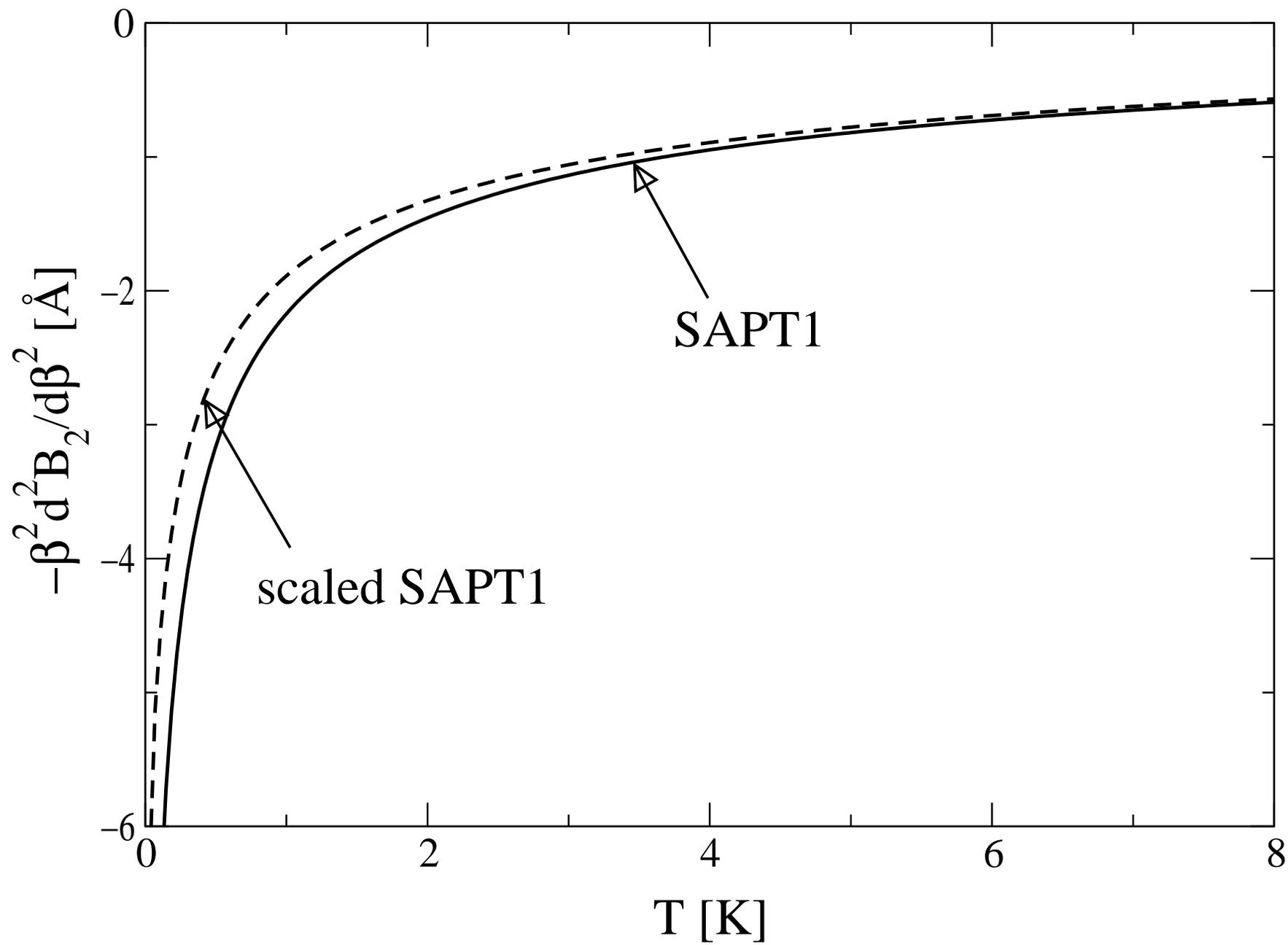
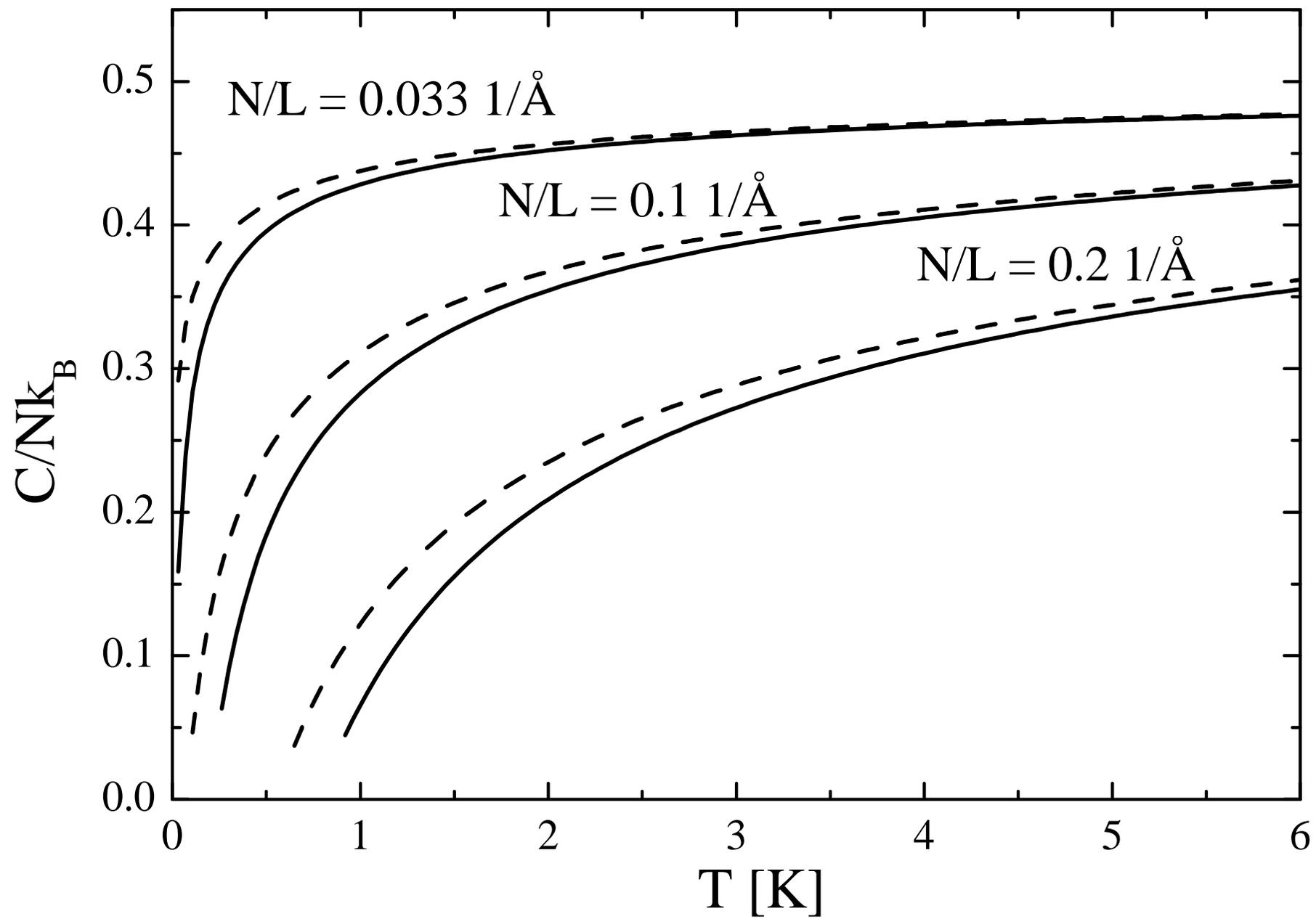
\end{document}